\documentclass[preprint2]{aastex}
\shorttitle{Polarimetry of HH 1--2 Region}
\shortauthors{Kwon et al.}

\newcommand{\mcoc}[1]{\multicolumn{1}{c}{{#1}}}

\usepackage{times}
\slugcomment{To appear in the Astrophysical Journal}

\begin{document}

\fontsize{10}{10.6}\selectfont

\title{Magnetic Field Structure of the HH 1--2 Region:
       Near-Infrared Polarimetry of Point-Like Sources}

\author{\sc Jungmi Kwon\altaffilmark{1,2},
            Minho Choi\altaffilmark{2}, Soojong Pak\altaffilmark{1,3},
            Ryo Kandori\altaffilmark{4}, Motohide Tamura\altaffilmark{4},
            Tetsuya Nagata\altaffilmark{5}, and Shuji Sato\altaffilmark{6}}
\affil{$^1$ Department of Astronomy and Space Science,
            Kyung Hee University, Yongin-si, Gyeonggi-do 446-701,
            South Korea}
\affil{$^2$ International Center for Astrophysics,
            Korea Astronomy and Space Science Institute,
            Daedukdaero 838, Yuseong, Daejeon 305-348, South Korea}
\affil{$^3$ soojong@khu.ac.kr}
\affil{$^4$ National Astronomical Observatory of Japan,
            Mitaka, Tokyo 181-8588, Japan}
\affil{$^5$ Department of Astronomy, Kyoto University, Kyoto 606-8502, Japan}
\affil{$^6$ Department of Astrophysics, Nagoya University,
            Nagoya 464-8602, Japan}

\begin{abstract}

\fontsize{10}{10.6}\selectfont

The HH 1--2 region in the L1641 molecular cloud was observed
in the near-IR $J$, $H$, and $K_s$ bands,
and imaging polarimetry was performed.
Seventy six point-like sources were detected in all three bands.
The near-IR polarizations of these sources
seem to be caused mostly by the dichroic extinction.
Using a color-color diagram,
reddened sources with little infrared excess were selected
to trace the magnetic field structure of the molecular cloud.
The mean polarization position angle of these sources is about 111\arcdeg,
which is interpreted as the projected direction of the magnetic field
in the observed region of the cloud.
The distribution of the polarization angle
has a dispersion of about 11\arcdeg,
which is smaller than what was measured in previous studies.
This small dispersion gives
a rough estimate of the strength of the magnetic field to be about 130 $\mu$G
and suggests that the global magnetic field in this region
is quite regular and straight.
In contrast, the outflows driven by young stellar objects in this region
seem to have no preferred orientation.
This discrepancy suggests
that the magnetic field in the L1641 molecular cloud
does not dictate the orientation of the protostars forming inside.
\end{abstract}


\section{INTRODUCTION}

Magnetic fields play a crucial role in various astrophysical processes,
including the evolution of interstellar molecular clouds and star formation
(Shu et al. 1987; Bergin \& Tafalla 2007; McKee \& Ostriker 2007).
One of the problems related to star formation
concerns the competition between magnetic and turbulent forces
(Mac Low \& Klessen 2004).
The magnetic field direction can be measured
by observing the dichroic polarization of background stars
in the optical and near-IR bands
and/or the linearly polarized emission from the dust grains
in the mid-IR and far-IR bands
(Davis \& Greenstein 1951; Matthews \& Wilson 2000).
The large-scale alignment of dust grains with the magnetic field
is known to be the cause of the dichroic extinction
and the interstellar polarization
seen in the direction of background sources.
Because of the low extinction,
near-IR imaging polarimetry is particularly useful
in tracing the dichroic polarization
of background stars and embedded sources seen through dense clouds
(Vrba et al. 1976; Wilking et al. 1979;
Tamura et al. 1987; Kandori et al. 2007).
Since both dichroic extinction and scattering processes
can contribute to the polarization of embedded sources,
multiwavelength polarimetry can be useful
in discriminating between the two mechanisms (Casali 1995).

The L1641 cloud is one of the nearest giant molecular clouds
and is a site of active star formation
(Kutner et al. 1977; Maddalena et al. 1986; Strom et al. 1989;
Morgan \& Bally 1991; Sakamoto et al. 1997; Zavagno et al. 1997).
The role of magnetic field in the star formation activity of L1641
is complicated.
With visual polarimetry of background stars,
Vrba et al. (1988) found
that the dispersion in position angles is large (33\arcdeg)
and suggested that the role of magnetic field in the global scale
is only incidental.
However, they also found
that the outflows in L1641 tend to be parallel to the field direction
and suggested that the role of magnetic field
is important in the local scale.
In contrast, Casali (1995) performed near-IR polarimetry
of young stellar objects (YSOs) in L1641
and found that the alignment of polarization vectors is poor,
which suggests that the magnetic field was not dominant
in the collapse dynamics.
To understand the situation better,
it was suggested that
more extensive polarimetry in the vicinity of each outflow and YSO
is necessary (Vrba et al. 1988).

One of the well studied parts of the L1641 cloud
is the region around the reflection nebula NGC 1999
and the Herbig-Haro objects HH 1--2.
This region contains several YSOs and outflows
(Herbig 1951; Haro 1952; Warren-Smith et al. 1980; Strom et al. 1989;
Corcoran \& Ray 1995; Choi \& Zhou 1997; Rodr{\'\i}guez et al. 2000).
The magnetic field structure in the HH 1--2 region has been studied
based on optical polarizations of point sources
(Strom et al. 1985; Warren-Smith \& Scarrott 1999, hereafter WS).
They found that the local magnetic field is directed
roughly along the axis of the HH 1--2 outflow,
but the number of detectable stars was too small
because of the large obscuration in this region.

In this paper, we present a wide-field near-IR polarimetry
of the HH 1--2 region.
In Section 2 we describe the observations and data reduction.
In Section 3 we present the results
of the polarimetry of point-like sources.
In Section 4 we discuss the magnetic field structure
and the star-forming activity in the HH 1--2 region.
A summary is given in Section 5.

\section{OBSERVATIONS}

\begin{figure*}
\epsscale{1.9}
\plotone{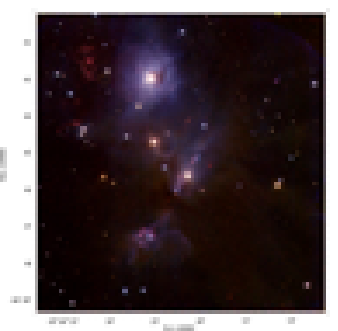}
\vspace{-0.5\baselineskip}
\centerline{\scriptsize [See http://minho.kasi.re.kr/Publications.html
for the original high-quality figure.]}
\vspace{-0.5\baselineskip}
\caption{
Color composite Stokes $I$ image of the HH 1--2 region
in the $J$({\it blue}), $H$({\it green}), and $K_s$({\it red}) bands
from the IRSF/SIRPOL observations.}
\end{figure*}

The observations toward the HH 1--2 region were carried out
using the SIRPOL imaging polarimeter
on the Infrared Survey Facility (IRSF) 1.4 m telescope
at the South African Astronomical Observatory.
SIRPOL consists of a single-beam polarimeter
(an achromatic half-wave plate rotator unit and a polarizer)
and an imaging camera (Nagayama et al. 2003).
The camera, SIRIUS, has three 1024 $\times$ 1024 HgCdTe infrared detectors.
IRSF/SIRPOL enables deep and wide-field
(7\farcm7 $\times$ 7\farcm7 with a scale of 0\farcs45 pixel$^{-1}$)
imaging polarimetry at the $J$, $H$, and $K_s$ bands simultaneously
(Kandori et al. 2006).

The observations were made on the night of 2008 January 9.
We performed 20 s exposures at 4 wave-plate angles
(in the sequence of 0\arcdeg, 45\arcdeg, 22$\fdg$5, and 67$\fdg$5)
at 10 dithered positions for each set.
The same observation sets were repeated 10 times
toward the target object and the sky backgrounds
for a better signal-to-noise ratio.
The total integration time was 2000 s per wave plate angle.
The typical seeing size during the observations
was $\sim$1\farcs3 in the $J$ band.
The polarization efficiencies of SIRPOL are stable over several years,
and the instrumental polarization is negligible (Kandori et al. 2006).
The efficiencies were measured
in 2007 December during a maintenance period,
just a few days before our observing run,
and were the same as the values reported by Kandori et al. (2006).

The data were processed using IRAF
in the same manner as described by Kandori et al. (2006),
which included dark-field subtraction, flat-field correction,
median sky subtraction, and frame registration.
Figure 1 shows the $J$-$H$-$K_s$ color composite intensity image
of the 8$'$ $\times$ 8$'$ region around HH 1--2
(hereafter the HH 1--2 field).
Many point-like sources in the HH 1--2 region were detected.
In addition to the point-like sources,
we detected HH 1--2, NGC 1999, and other nebulosities,
but the polarimetric studies of these extended sources
will be presented elsewhere in the future.

\begin{figure*}
\epsscale{1.9}
\plotone{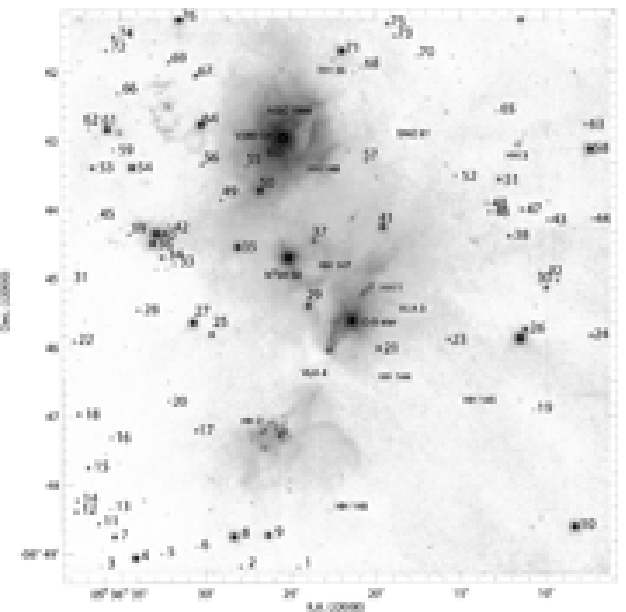}
\vspace{-0.5\baselineskip}
\centerline{\scriptsize [See http://minho.kasi.re.kr/Publications.html
for the original high-quality figure.]}
\vspace{-0.5\baselineskip}
\caption{
Finding chart of the HH 1--2 field (color-negative image of Fig. 1).
Detected point-like sources are labeled (Table 1).
Bright stars and some extended sources are also labeled.}
\end{figure*}

\section{RESULTS}

\subsection{Photometry}

The IRAF DAOPHOT package was used for source detection and photometry
(Stetson 1987).
The DAOPHOT program automatically detected
point-like sources with peak intensities
greater than 10$\sigma$ above the local sky background,
where $\sigma$ is the rms uncertainty.
The automatic detection procedure
misidentified some spurious sources and missed some real sources,
and the source list was corrected
by visually inspecting the images carefully.
The pixel coordinates of the detected sources
were matched with the celestial coordinates of their counterparts
in the Two Micron All Sky Survey (2MASS) Point Source Catalog.
The IRAF IMCOORDS package was applied to the matched list
to obtain plate transform parameters.
The rms uncertainty in the coordinate transformation was $\sim$0\farcs1.

Aperture photometry was performed again with the resulting images.
The aperture radius was 3 pixels,
and the sky annulus was set to 10 pixels with a 5 pixel width.
The resulting list contains 76 sources
whose photometric uncertainties
are less than 0.1 mag in all three bands (Table 1).
These point-like sources are labeled in Figure 2.
Four bright sources (V380 Ori, the C-S star, N$^3$SK 50,
and an unnamed star $\sim$10$''$ south of source 26) were saturated,
and they were excluded from the list.

The Stokes $I$ intensity of each point-like source was calculated by
\begin{equation}
   I = {1\over2} (I_{0} + I_{22.5} + I_{45} + I_{67.5}),
\end{equation}
where $I_a$ is the intensity with the half wave plate oriented at $a$\arcdeg.
The magnitude and color of the photometry
were transformed into the 2MASS system by
\begin{equation}
   \rm MAG_{2MASS} = MAG_{IRSF} + \alpha_1 \times COLOR_{IRSF} + \beta_1
\end{equation}
and
\begin{equation}
   \rm COLOR_{2MASS} = \alpha_2 \times COLOR_{IRSF} + \beta_2,
\end{equation}
where MAG$_{\rm IRSF}$ is the instrumental magnitude from the IRSF images,
and MAG$_{\rm 2MASS}$ is the magnitude from the 2MASS Point Source Catalog.
The parameters were determined by fitting the data
using a robust least absolute deviation method.
For the magnitudes,
$\alpha_1$ = 0.017, --0.064, and 0.001,
and $\beta_1$ = --4.986, --4.717, and --5.375
for $J$, $H$, and $K_s$, respectively.
For the colors,
$\alpha_2$ = 1.007 and 0.960, and $\beta_2$ = --0.261 and 0.664
for $J - H$ and $H - K_s$, respectively.
The coefficients $\beta_1$ and $\beta_2$
include both the zero point correction and aperture correction.
The derived magnitudes are listed in Table 1.
The 10$\sigma$ limiting magnitudes
were 19.6, 18.7, and 17.3 for $J$, $H$, and $K_s$, respectively.

\subsection{Polarimetry}

Aperture polarimetry was carried out
on the combined intensity images for each wave plate angle,
instead of using the Stokes $Q$ and $U$ images.
This is because the center of the sources
cannot be determined satisfactorily on the $Q$ and $U$ images.
From the aperture photometries on each wave plate angle image,
the Stokes parameters of each point-like source
were derived by
\begin{equation}
   Q = I_{0} - I_{45}
\end{equation}
and
\begin{equation}
   U = I_{22.5} - I_{67.5}.
\end{equation}
The aperture and sky radius
were the same as those used in the photometry of $I$ images.
The degree of polarization, $P$,
and the polarization position angle, $\theta$,
can be calculated by
\begin{equation}
   P_0 = {{\sqrt{Q^2 + U^2}} \over {I}},
\end{equation}
\begin{equation}
   P = \sqrt {P_0^2 - \delta P^2},
\end{equation}
and
\begin{equation}
   \theta = {1\over2} \arctan {U \over Q},
\end{equation}
where $\delta P$ is the uncertainty in $P_0$.
Equation (7) is necessary to debias the polarization degree
(Wardle \& Kronberg 1974).
Finally, $P$ was corrected using the polarization efficiencies of SIRPOL:
95.5 \%, 96.3 \%, and 98.5 \% at $J$, $H$, and $K_s$, respectively
(Kandori et al. 2006).

\begin{figure}[!t]
\epsscale{1.0}
\plotone{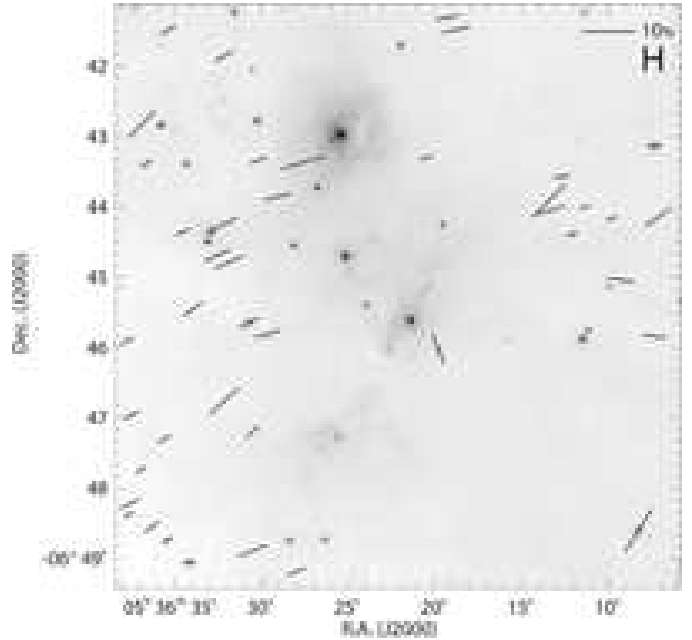}
\vspace{-0.5\baselineskip}
\centerline{\scriptsize [See http://minho.kasi.re.kr/Publications.html
for the original high-quality figure.]}
\vspace{-0.5\baselineskip}
\caption{
Stokes $I$ image of the $J$ band with polarization vectors.
The length of the vectors is proportional to the degree of polarization.
Shown in the upper right corner is a 10\%\ vector.
Note that there are bad pixel clusters
around the upper-left and upper-right corners
and the middle of the right boundary.}
\end{figure}

\begin{figure}[!t]
\epsscale{1.0}
\plotone{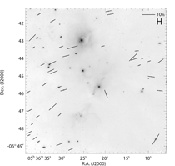}
\vspace{-0.5\baselineskip}
\centerline{\scriptsize [See http://minho.kasi.re.kr/Publications.html
for the original high-quality figure.]}
\vspace{-0.5\baselineskip}
\caption{
The same as Fig. 3 for the $H$ band.}
\end{figure}

\begin{figure}[!t]
\epsscale{1.0}
\plotone{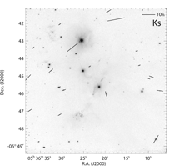}
\vspace{-0.5\baselineskip}
\centerline{\scriptsize [See http://minho.kasi.re.kr/Publications.html
for the original high-quality figure.]}
\vspace{-0.5\baselineskip}
\caption{
The same as Fig. 3 for the $K_s$ band.}
\end{figure}

\begin{figure}[!t]
\epsscale{1.0}
\plotone{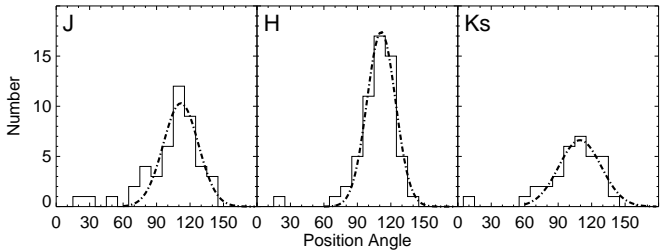}
\caption{
Histograms of polarization position angles
for the $J$, $H$, and $K_s$ bands.
All the sources in Table 2 with detected polarization are included.
Dot-dashed curves: Gaussian fits.
The peak angle and dispersion
are 112\arcdeg\ and 16\arcdeg\ for $J$,
111\arcdeg\ and 13\arcdeg\ for $H$,
and 110\arcdeg\ and 19\arcdeg\ for $K_s$.}
\end{figure}

Table 2 shows the derived source parameters.
The uncertainties given in Table 2 (and elsewhere in this paper)
are 1$\sigma$ values.
Figures 3--5 show the polarization vector maps of point-like sources
superposed on the $I$ images.
For the sources with $P$/$\delta P$ $\geq$ 4 and $P <$ 9\% (21 sources),
the correlation coefficients are 0.91 for ($\theta_H$, $\theta_{K_s}$)
and 0.97 for ($\theta_H$, $\theta_J$).
Figure 6 shows the histograms and Gaussian fits
for the polarization position angles.
Each of the three histograms shows a single peak at $\sim$111\arcdeg.
Note that the dispersion in $\theta$ is smallest in the $H$ band.
In addition, for most sources,
the signal-to-noise ratio ($P$/$\delta P$) is higher
in the $H$ band than the other bands.
The $H$ band polarimetry is more reliable than those of the other bands
probably because the contamination from extended nebulosity
is smaller in the $H$ band than in the $J$ band
and because the dichroic polarization
is more efficient in the $H$ band than in the $K_s$ band.
Therefore, our discussion in Section 4
will be mainly based on the $H$ band data.

The relation between the polarimetric and spectral data
may be useful in understanding the nature of polarization.
The degree of polarization
appears to be correlated with near-IR colors (Fig. 7).
The empirical relation for the upper limit of interstellar polarization
suggested by Jones (1989) is
\begin{equation}
   P_{K, \rm max} = \tanh \left\{ 1.5 E(H-K) {{1-\eta}\over{1+\eta}} \right\},
\end{equation}
where $\eta$ = 0.875 and $E(H-K)$ is the reddening owing to extinction.
Most of the sources are within this limit (Fig. 7).
A few sources are above the $P_{\rm max}$ limit,
but their uncertainties are large.
The near-IR polarization-to-extinction efficiency
of the point-like sources in the HH 1--2 field
is consistent with that caused by aligned dust grains
in the dense interstellar medium.
Therefore, their polarizations are likely dominated
by the interstellar dichroic extinction,
and the intrinsic polarization, if any,
did not significantly enhance the degree of polarization.
However, this result does not completely exclude the possibility
that some of the sources have intrinsic polarization
because depolarization is also possible.

\begin{figure}[!t]
\epsscale{1.0}
\plotone{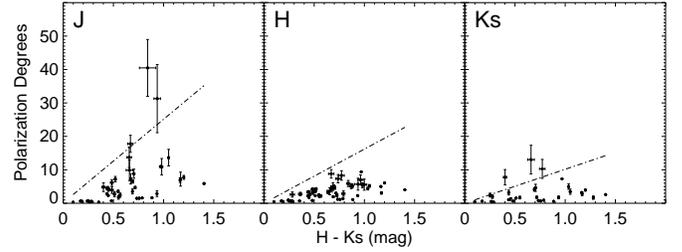}
\caption{
Degree of polarization vs. $H - K_s$ color.
Dot-dashed lines:
empirical upper limits ($P_{\rm max}$; Jones 1989).}
\end{figure}

\section{DISCUSSION}

\subsection{Comparison with Previous Studies}

Polarimetry of bright point-like sources in the HH 1--2 region
was reported previously by several authors.
These studies covered larger regions than our study,
but they were much shallower.
Strom et al. (1985) carried out $I$ band polarimetry
and measured the polarization position angle of two sources:
157\arcdeg\ for the C-S star and 135\arcdeg\ for N$^3$SK 50.
Casali (1995) measured the $K$ band polarization angle of two sources:
105\arcdeg\ for V380 Ori and 154\arcdeg\ for N$^3$SK 50.
These three sources were saturated in our observations,
and no direct comparison is possible.

WS presented a broad-band (450--1000 nm) polarimetry
of eight point-like sources in a larger ($\sim$10$'$) region.
The polarization position angle averaged about 130\arcdeg\
with a dispersion of 30\arcdeg.
The polarization angle of bright sources were
125\arcdeg\ for the C-S star and 128\arcdeg\ for N$^3$SK 50.
Direct comparisons for polarization angles are possible for three sources,
while meaningful comparisons for polarization degrees
are difficult due to wavelength dependence.
WS 1 (source 1 of WS) corresponds to our source 8,
and the polarization position angle of WS (145\arcdeg\ $\pm$ 5\arcdeg)
is somewhat larger than our near-IR measurements (95--121\arcdeg).
WS 5 and WS 6 correspond to our sources 10 and 58, respectively,
and their polarization orientations agree reasonably well, within 5\arcdeg.
It is not clear what caused the difference of source 8.
One of the possibilities is
that the polarization of source 8 is caused
mostly by the scattering process rather than dichroic extinction,
as this source shows little extinction
(see Section 4.7 for more discussions).

Based on the broad-band polarimetry,
WS suggested that the magnetic field in the HH 1--2 region
is oriented at a position angle of 126\arcdeg.
This interpretation, however, should be corroborated by deeper observations
because the size of their source sample was too small.
Our observations can provide statistically more significant interpretations,
which are presented below.

\subsection{Source Classification}

To study the magnetic field structure of molecular clouds
through the interstellar polarization caused by dichroic extinction,
it is necessary to select sources without intrinsic polarization.
YSOs in the cloud can exhibit a substantial degree of intrinsic polarization
caused by circumstellar material.
Such sources may show a large amount of infrared excess emission.
Therefore, it is important to classify sources,
and the multiwavelength photometry can be useful.

\begin{figure}
\epsscale{0.9}
\plotone{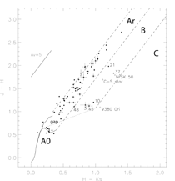}
\vspace{-0.5\baselineskip}
\centerline{\scriptsize [See http://minho.kasi.re.kr/Publications.html
for the original high-quality figure.]}
\vspace{-0.5\baselineskip}
\caption{
Color-color diagram of the point-like sources in the HH 1--2 field.
Filled circles:
point-like sources from this work (Table 1).
Open circles:
bright sources (Strom et al. 1989; Carpenter et al. 2001).
Solid curve:
locus of main sequence and giant branch stars (Bessell \& Brett 1988).
Dotted line:
locus of classical T Tauri stars (Meyer et al. 1997).
Dashed lines:
boundaries between domains Ar, B, and C (see Section 4.2).
Solid line:
reddening vector.}
\end{figure}

Figure 8 shows a color-color diagram
for all the sources detected in all three bands.
The diagram was divided into several domains.
Based on the location in this diagram,
sources can be classified into a few groups (Lada \& Adams 1992).
The area near the locus of main-sequence/giant stars is called domain A0.
Sources in domain A0 are either field stars (dwarfs and giants)
or pre-main-sequence (PMS) stars with little infrared excess
(weak-lined T Tauri stars and some classical T Tauri stars)
and with little reddening.
There is a clear gap just above domain A0,
and the area above this gap in the direction of the reddening vector
is called domain Ar.
Sources in domain Ar are either field stars or PMS stars
with little infrared excess and with substantial reddening.
Domain B is the area next to domain Ar
in the direction of higher $H - K_s$ (to the right)
and above the locus of classical T Tauri stars.
Sources in domain B are PMS stars with infrared excess emission from disks.
Domain C is the area next to domain B to the right.
Sources in domain C are infrared protostars or Class I sources.
Herbig AeBe stars tend to occupy lower parts of domains B and C.
This classification based on the color-color diagram, however,
is far from perfect.
A certain fraction of classical T Tauri stars may reside in domains A0 or Ar,
some protostars can be found in domain B,
and some extremely reddened AeBe stars may be found among protostars
(Lada \& Adams 1992).
This ``contamination'' will eventually contribute to the uncertainty
in statistical quantities derived from the classification,
but the estimation of this uncertainty is beyond the scope of this paper.

There are thirteen sources in domain A0,
and they are collectively called group A0.
They are either foreground stars or
those seen along lines of sight with little extinction.

Fifty seven sources were found in domain Ar.
These sources (group Ar) are either background stars
or PMS stars in the L1641 cloud.
They are the most useful sources
for the study of the magnetic fields in the cloud (Section 4.4).

Six sources were found in domain B.
These sources (group B) may be PMS stars associated with the L1641 cloud.
Source 10 (WS 5) is the emission-line star AY Ori (Wouterloot \& Brand 1992)
and also source 13932 of Carpenter et al. (2001).
Source 21 is source 14843 of Carpenter et al. (2001).
In addition, two of the brightest objects in this field,
the C-S star and N$^3$SK 50, also belong to group B.

None of our sources are located in domain C.
However, this nondetection does not mean
that there is no protostar in the HH 1--2 field.
Some protostars (for example, HH 1--2 VLA 1) are deeply embedded
and undetectable in the near-IR bands.
In addition, the red protostars tend to be missed
at shorter wavelengths ($J$ and/or $H$ bands).
The Herbig Ae star V380 Ori
is located in the lower part of domain C, as expected.

\subsection{Source of Polarization}

To study the magnetic field structure of L1641,
we will use the distribution of the polarization of group Ar sources
(Section 4.4).
To do that, the source of polarization needs some discussion.
The distribution of polarization angle (Fig. 6)
shows a well-defined single peak,
which suggests that the cause of polarization is relatively simple.
Especially for the group Ar source,
the polarization seems to be mostly caused by the L1641 cloud,
based on several lines of evidence described below.

First, optically thick molecular lines (such as CO and H$_2$CO)
observed toward the HH 1--2 region
show a single velocity component at $\sim$8 km s$^{-1}$
(Loren et al. 1979; Snell \& Edwards 1982),
which suggests that L1641 is the only molecular cloud
that is optically thick enough to cause the dichroic extinction.
The spectrum of $^{13}$CO, however, shows
that there are two velocity components:
a stronger one at $\sim$8.5 km s$^{-1}$
and a weaker one at $\sim$7 km s$^{-1}$ (Edwards \& Snell 1984),
and both components belong to the L1641 cloud (Sakamoto et al. 1997).
The stronger component is directly related to the dense gas
in the HH 1--2 region,
and the weaker component is probably related to the dense gas
with a column density peak located $\sim$20$'$ north of HH 1--2
(Takaba et al. 1986).
Since the distribution of the polarization angle
shows a well-defined single peak,
one of them (most likely the $\sim$8.5 km s$^{-1}$ component)
probably dominates in the polarization process.
Alternatively, the magnetic field direction of the two components
may be very similar.

\begin{figure}[!t]
\epsscale{1}
\plotone{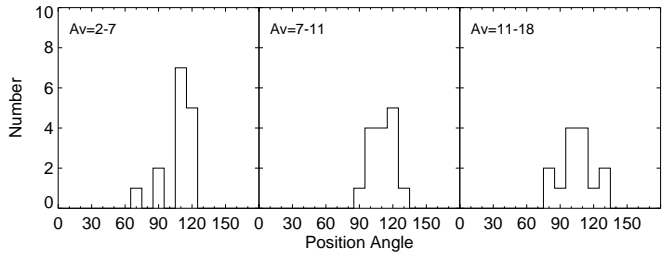}
\caption{
Histograms of the $H$ band polarization position angles of group Ar.
The sources were grouped by the amount of extinction:
$A_V$ = 2--7, 7--11, and 11--18.
The subgroups contain 15, 15, and 14 sources, respectively.}
\end{figure}

Second, the peak polarization angle
is insensitive to the amount of extinction.
If group Ar sources would have a preferred direction of polarization
even before their IR photons are affected by the L1641 cloud,
sources with a small extinction would show
the effect of this ``background'' polarization direction,
while sources with a large extinction
would reflect only the polarization caused by L1641.
Figure 9 shows the histograms of the polarization position angle
for three subgroups of sources grouped by the amount of extinction.
All the three subgroups show a peak angle at $\sim$110\arcdeg.
Therefore, the polarization of the group Ar sources
are mostly caused by the L1641 cloud only,
and the distribution of the polarization angle
may be a very good tracer of the magnetic field structure of L1641.

Third, the Galactic latitude of L1641 is high ($b$ = --19\fdg8).
It is unlikely that light from distant luminous giants
would suffer a significant amount of dichroic extinction
by any other cloud behind L1641.

For some clouds, the polarization direction of background stars
can be caused by several sources of polarization along the line of sight.
For example, in the direction
of the Southern Coalsack dark cloud ($b \approx$ --1\arcdeg),
there are at least three components of polarization
(Andersson \& Potter 2005).
In such cases, the interpretation of the polarization
and its relation with the magnetic field structure
can be quite complicated.
In the case of the HH 1--2 region,
the polarization seems to be mostly caused by L1641,
and the relation between the polarization direction
and the magnetic field structure is relatively straightforward,
as discussed in the next section.

\subsection{Magnetic Field Structure}

The sources in group Ar are best
for studying the magnetic field structure of the molecular cloud
because they are subject to dichroic extinction
and because they would have relatively little intrinsic polarization.
Figure 10 shows the histogram and Gaussian fit
for the polarization position angles of group Ar sources.
The peak angle is 111\arcdeg, and the dispersion is 11\arcdeg.
Comparing Figures 6 and 10,
it is clear that selecting only group Ar sources
makes the statistical noise smaller:
outliers disappeared, and the dispersion became smaller.
Therefore, we suggest that the global magnetic field in the HH 1--2 region
is oriented at a position angle of $\sim$111\arcdeg.
This orientation is consistent with the large-scale field structure of L1641
(Vrba et al. 1988).
Previous studies of the HH 1--2 region (Strom et al. 1985; WS)
suggested larger position angles and larger dispersions
because their sample sizes were too small
and because they included bright PMS stars in the sample.

\begin{figure}[!t]
\epsscale{1.0}
\plotone{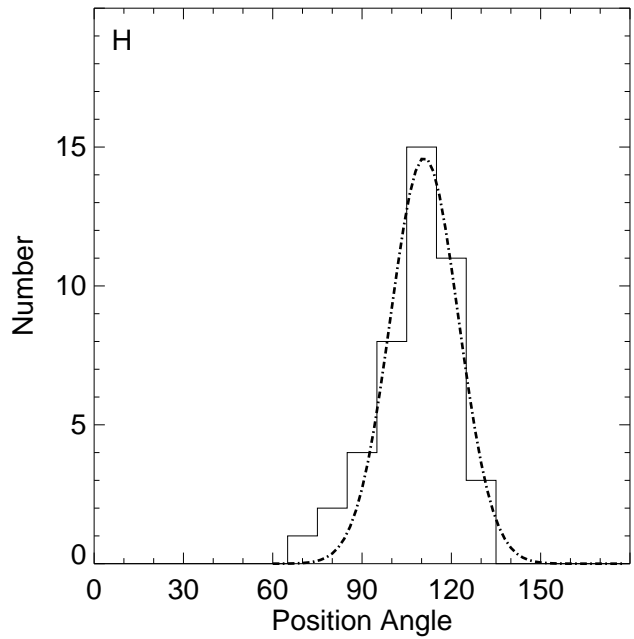}
\caption{
Histogram of polarization position angles
for the group Ar sources in the $H$ band.
Dot-dashed curve: Gaussian fit.}
\end{figure}

\begin{figure}[!t]
\epsscale{1}
\plotone{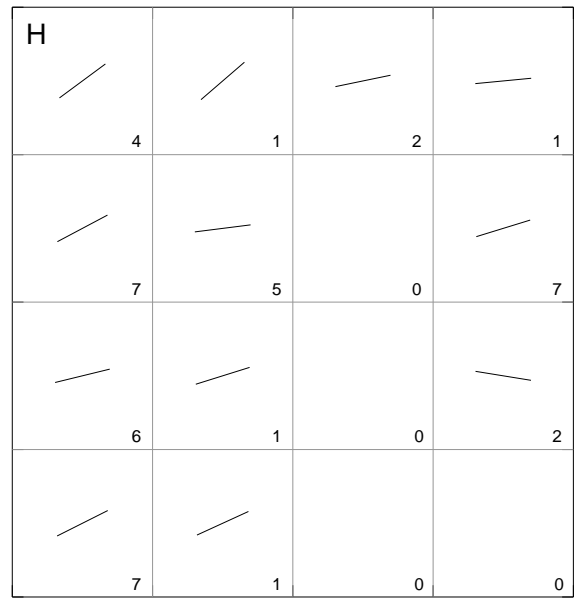}
\caption{
Average polarization position angle of group Ar sources
in 2$'$ $\times$ 2$'$ subregions of the HH 1--2 field.
The number of sources in each subregion is labeled.}
\end{figure}

To see whether there is a systematic gradient
of magnetic field orientation over the imaged field,
we divided the HH 1--2 field into 16 (4 $\times$ 4) subregions.
In each subregion, group Ar sources were selected,
and their polarization position angles were averaged.
Inspection of the resulting distribution of polarization angle (Fig. 11)
did not reveal any systematic trend.
Therefore, we suggest
that the measured dispersion of 11\arcdeg\ represents
the local variation of magnetic field orientation.
Here, ``local'' means each line of sight,
though it should be noted that the variation along the line of sight
would affect the polarization in an integrative way.

Although the polarization measurement
does not provide a direct estimate of the magnetic field strength
at each data point in the image,
a rough estimation over a large region is possible
by statistically comparing the dispersion of polarization orientation
with the degree of turbulence in the cloud
(Chandrasekhar \& Fermi 1953).
Assuming that velocity perturbations are isotropic,
the strength of the magnetic field projected on the plane of the sky
can be calculated by
\begin{equation}
   B_p = {\cal Q} \sqrt{4\pi\rho}\ {{\delta v_{\rm los}}\over{\delta\theta}},
\end{equation}
where ${\cal Q}$ is a factor to account for various averaging effects,
$\rho$ is the mean density of the cloud,
$\delta v_{\rm los}$ is the rms line-of-sight velocity,
and $\delta\theta$ is the dispersion of polarization angles.
Ostriker et al. (2001) suggested that ${\cal Q} \approx 0.5$
is a good approximation
when the angle dispersion is small ($\delta\theta \lesssim$ 25\arcdeg)
from numerical simulations.
From the observations of the molecular condensation
in the CO and $^{13}$CO $J$ = 1 $\rightarrow$ 0 lines,
Takaba et al. (1986) estimated
an H$_2$ column density of 2.5 $\times$ 10$^{22}$ cm$^{-2}$
and a size of 2.0 pc.
The density $\rho$ can be derived
by assuming that the line-of-sight size of the dense condensation
is similar to the lateral size.
The FWHM line width of the C$^{18}$O $J$ = 1 $\rightarrow$ 0 line,
2.7 km s$^{-1}$ (Takaba et al. 1986),
can be used to estimate $\delta v_{\rm los}$.
Then the derived field strength is $B_p \approx$ 130 $\mu$G.
The uncertainty in this estimate may be rather large
because the observed HH 1--2 field is only a part of the L1641 cloud,
and it should be taken as an order-of-magnitude estimate.
The estimated magnetic field strength of the HH 1--2 region
is similar to that of other molecular clouds (20--200 $\mu$G)
derived using the Chandrasekhar-Fermi method
(e.g., Andersson \& Potter 2005; Poidevin \& Bastien 2006;
Alves et al. 2008).

An interesting issue is how good near-IR polarimetry is
in tracing the magnetic field structure of dense clouds.
Goodman et al. (1995) suggested
that the polarizing power of dust grains may drop
in the dense interior of some dark clouds
and that near-IR polarization maps of background sources may be unreliable.
However, the relevant physics is surprisingly complex (Lazarian 2007),
and there are observational evidence and theoretical explanations
for aligned grains in dense cloud cores
(Ward-Thompson et al. 2000; Cho \& Lazarian 2005).
In the case of our study of the HH 1--2 field,
the polarization degree $P_H$ does not show a clear sign of saturation
up to $H - K_s \approx 1$ (Fig. 7),
and the observed polarizations of detected sources
may well represent the magnetic fields in the HH 1--2 region.

\subsection{Pre-Main-Sequence Stars with Infrared Excess}

Sources in group B are PMS stars with infrared excess.
In the study of global magnetic fields,
we excluded group B sources
because they may have intrinsic polarizations.
To verify this precaution,
their polarization properties may be compared with those of group Ar.

\begin{figure}
\epsscale{1}
\plotone{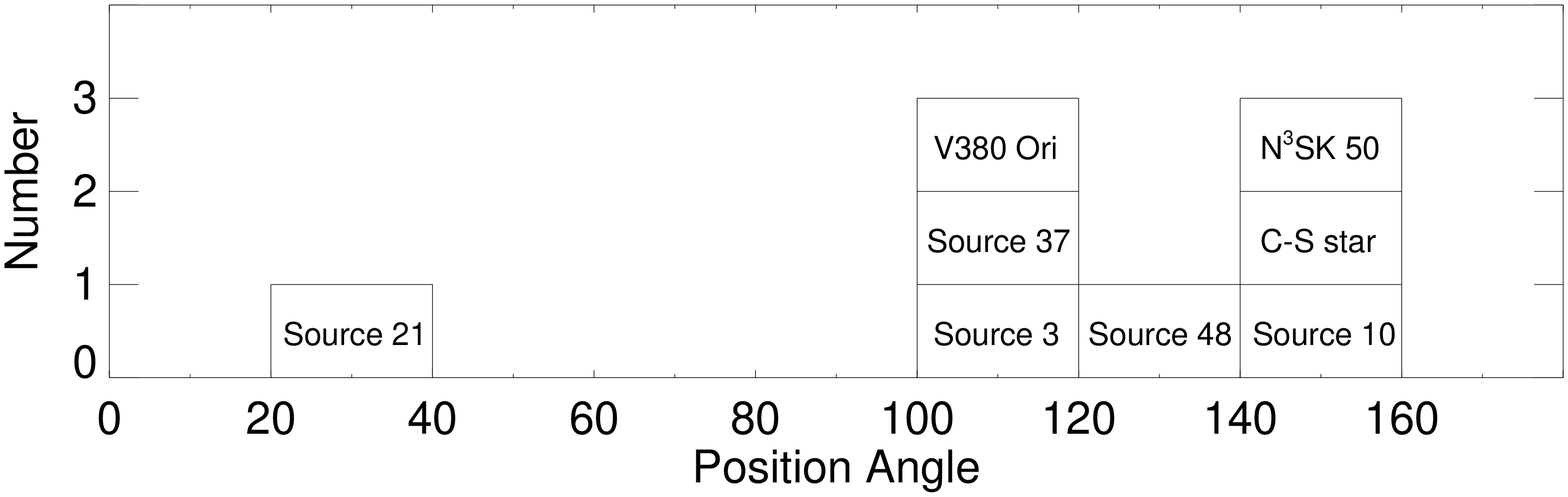}
\caption{
Histogram of near-IR polarization position angles
for the group B sources and the bright PMS stars
(Strom et al. 1985; Casali 1995).
Note that a 20\arcdeg\ bin is used because the sample size is small.}
\end{figure}

Figure 12 shows the histogram of the polarization position angles
of group B sources from this work
and three bright PMS stars from the literature.
The distribution is relatively widespread,
and the peak angle is ill-defined.
Probably it is not a single-peaked distribution (cf. Tamura \& Sato 1989),
but the sample size is too small to discuss in detail.
Quantifying the dispersion is difficult,
but it is clearly much larger than that of group Ar.
This distribution implies
that these PMS stars have significant intrinsic polarizations
and that the orientation of the intrinsic polarization
has no strong correlation with that of the global magnetic fields.

\subsection{Outflow Orientations}

The alignment between the magnetic fields of a star-forming cloud
and the orientation of resulting stars is an interesting topic,
because it can provide information on the role of magnetic fields
during the protostellar collapse and subsequent evolution of the system.
In an ideal scenario of quasi-static isolated single star formation,
one would expect
that the symmetry axis of the newly formed (proto)stellar system
(such as the rotation axis, bipolar outflow axis)
would be parallel to the magnetic field
that threaded the initial dense cloud core (e.g., Galli \& Shu 1993),
which would imply
that the distribution of outflow orientations
would be similar to that of the polarization angle of background sources
(with a correction for the averaging effects, see below).
Strom et al. (1986) found
that about 70\% of the optical flows in their imaging survey
have the outflow directions within 30\arcdeg\
of the magnetic field direction of the cloud they belong to.
However, in a recent survey of classical T Tauri stars,
M{\'e}nard \& Duch{\^e}ne (2004) found
that the jets/disks around T Tauri stars are oriented randomly
with respect to the cloud magnetic fields.

In comparing the distributions
of the polarization angle and the outflow direction,
the dispersion of the polarization angle should be corrected
by a certain factor,
because the averaging effects along the line of sight
would decrease the dispersion.
While it is impossible to find an exact correction factor for each cloud,
numerical simulations suggests that it is in the range of 0.46--0.51
when the dispersion is small (Ostriker et al. 2001).
Andersson \& Potter (2005) tackled this problem using Monte Carlo simulations
to analyze the polarimetry of the Southern Coalsack cloud
and found that a reasonable correction factor
would be in the range 0.3--1.0.
This factor tends to be large (close to 1)
when the cloud is nearly homogeneous
and can be small when there are many distinct regions
along the line of sight.
The HH 1--2 region is not completely homogeneous,
as the molecular cloud has two velocity components along the line of sight
(see Section 4.3),
and also not as complicated as the Southern Coalsack cloud.
Therefore, a correction factor of $\sim$0.5 would be a reasonable value.

For the HH 1--2 region,
previously much attention was paid to the relation
between the direction of the HH 1--2 outflow
(position angle $\approx$ 148\arcdeg)
and the orientation of global magnetic fields (Strom et al. 1985; WS).
Our polarimetry shows
that the difference between them ($\sim$40\arcdeg)
is much larger than the 11\arcdeg\ dispersion of the polarization angles.
Therefore, there is a certain degree of misalignment.
To make a more meaningful comparison,
a list of optical jets in the HH 1--2 field with known flow directions
was compiled (Table 3).
The driving sources of these outflows are YSOs.
Figure 13 shows the distribution of the outflow orientations.
Considering the distribution of the polarization angle
and the correction factor mentioned above,
the expected distribution of the outflow direction
would show a single peak at $\sim$110\arcdeg\
with a dispersion of $\sim$20\arcdeg.
Surprisingly, the position angle of the outflow shown in Figure 13
seems to be nearly random,
i.e., the distribution is almost uniform,
which makes a stark contrast
to the distribution of polarization angles.
Therefore, we suggest that many protostars
may become disoriented during the star forming process.

\begin{figure}[!t]
\epsscale{1}
\plotone{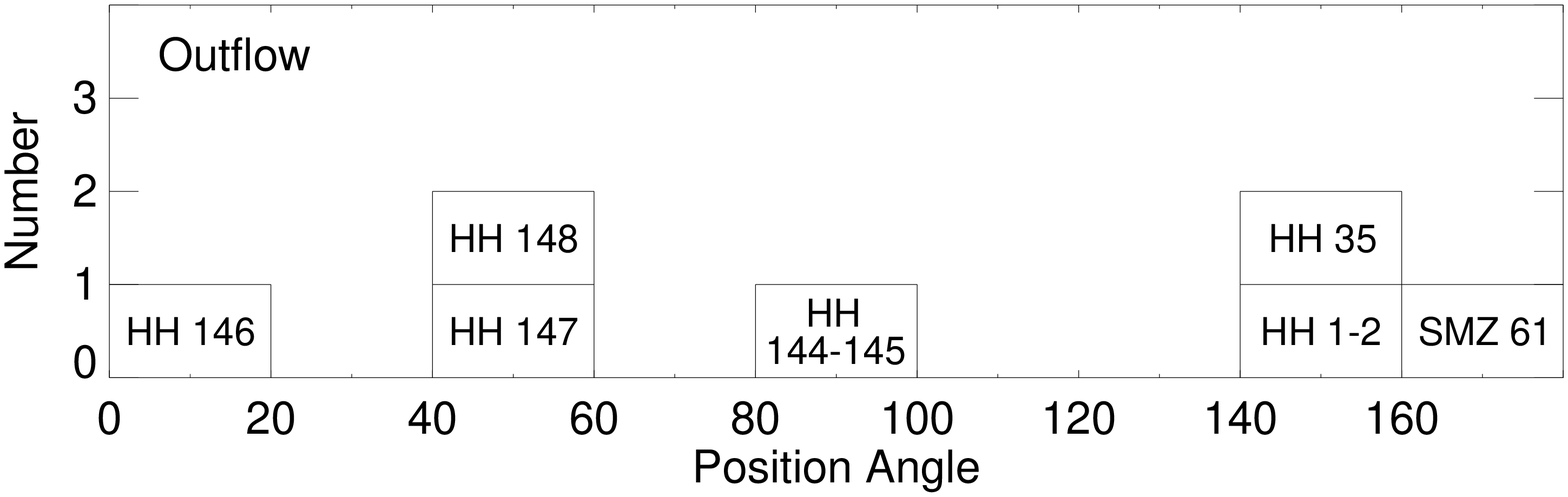}
\caption{
Histogram of outflow orientations of optical jets in the HH 1--2 region.}
\end{figure}

The random orientation of YSOs suggests
that the global magnetic field
may lose control of a collapsing protostellar core
at a certain small scale or at a certain stage of evolution.
Several mechanisms may be working to cause the disorientation.
First, the dynamical interaction
during a binary (or multiple star) formation process
can produce a misaligned system (e.g., Bonnell et al. 1992).
In fact, misaligned binaries in the Class II phase are not unusual
(Monin et al. 2007).
Among the outflow driving sources in the HH 1--2 region,
V380 Ori and HH 1--2 VLA 1/2 are examples to the point.
V380 Ori is a 0\farcs15 ($\sim$60 AU) binary and drives HH 35 and HH 148,
and these two outflows are almost perpendicular to each other
(Strom et al. 1986; Leinert et al. 1997).
HH 1--2 VLA 1/2 is a 3$''$ ($\sim$1200 AU) binary
and drives HH 1--2 and HH 144--145,
and the projected angle between these flows is $\sim$70\arcdeg\
(Reipurth et al. 1993).
Second, even in a single-star formation,
an outflow may not be aligned
with the magnetic field of the surrounding cloud
if the magnetic field is too weak (Matsumoto et al. 2006).
There are other possible mechanisms (M{\'e}nard \& Duch{\^e}ne 2004).

Considering that there are Class 0 sources showing misalignments,
the disorientation may happen very early.
For example, HH 1--2 VLA 1 is a Class 0 source (Chini et al. 1997).
Examples in other star-forming regions include
HH 24 MMS, which seems to have two accretion disks
with a $\sim$45\arcdeg\ difference in projected orientation
(Kang et al. 2008),
and NGC 1333 IRAS 2, which seems to drive two outflows
almost perpendicular to each other (Hodapp \& Ladd 1995).

\subsection{Wavelength Dependence of Interstellar Polarization}

Though the selected point-like sources have no detectable nebulosity,
we cannot rule out the possibility of the presence
of unresolved reflection nebulae in some of the sources.
Indeed, Casali (1995) showed
that the polarization by scattering is important
in some of the sources in L1641,
especially the sources with low extinction.
To discriminate between the contributions
from the dichroic extinction and from the scattering,
the wavelength dependence of polarization can be measured
by calculating the ratio of polarization degrees.
In the infrared wavelength range,
the polarization by dichroic extinction decreases with wavelength
while the polarization by scattering is not a strong function of wavelength
(Whittet et al. 1992; Casali 1995).

\begin{figure}[!t]
\epsscale{1.0}
\plotone{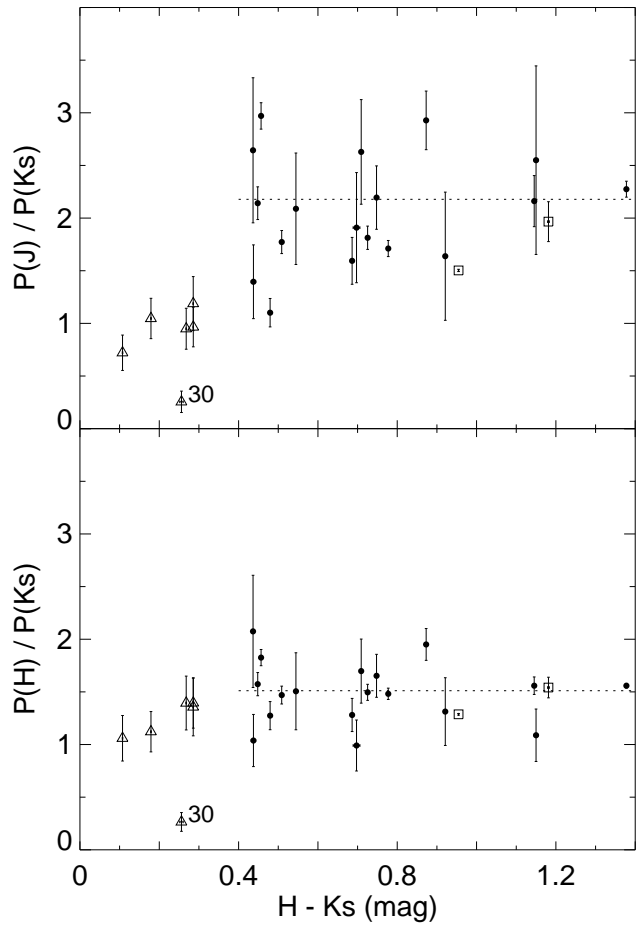}
\caption{
The ratio of $J$ to $K_s$ polarizations ({\it top panel})
and the ratio of $H$ to $K_s$ polarizations ({\it bottom panel})
against the $H - K_s$ color
for the 26 sources detected in the $J$, $H$, and $K_s$ bands.
Triangles:
sources in group A0.
Source 30 has unusually low ratios.
Filled circles:
sources in group Ar.
Squares:
sources in group B.
Dotted lines:
weighted average for group Ar.}
\end{figure}

Figure 14 shows the $P_J/P_{K_s}$ and $P_H/P_{K_s}$ ratios
for the sources in the HH 1--2 field.
It is very clear that groups A0 and Ar show very different behavior.
Within group Ar, the ratios are reasonably constant,
and the weighted average values are
$P_J/P_{K_s}$ = 2.18 and $P_H/P_{K_s}$ = 1.51
with standard deviations of 0.10 and 0.07, respectively.
These values are consistent with those for the whole L1641 area
(2.5 and 1.4 with 2$\sigma$ scatters of 0.7 and 0.2, respectively;
Casali 1995),
which is also consistent with the empirical relation
$P \propto \lambda^{-\beta}$ with $\beta$ = 1.6--2.0 (Whittet 1992).
Therefore, the polarization of group Ar sources
can be very well explained by dichroic extinction.

In contrast, the $P_J/P_{K_s}$ and $P_H/P_{K_s}$ ratios
of group A0 sources are near unity,
which is significantly different from those of group Ar sources.
Therefore, the polarization mechanism for low extinction sources
may be dominated by the circumstellar scattering
(see more discussions in Section 5.4 of Casali 1995).
Source 30 shows unusually low ratios (Fig. 14)
because its $K_s$ band polarization degree is high
while the $P_J/P_H$ ratio is near one as expected.
Its $K_s$ band polarization angle is
also quite different from those of $J$ and $H$ bands.
It is not clear what caused this peculiar behavior.

\section{SUMMARY}

We conducted a deep and wide-field $J$-$H$-$K_s$ imaging polarimetry
toward a 8$\arcmin$ $\times$ 8$\arcmin$ region around HH 1--2
in the star-forming cloud L1641.
The main results in this study are summarized as follows.

\begin{enumerate}

\item
Aperture photometry of point-like sources in the HH 1--2 field was made.
The number of sources detected in all three bands is 76.
These sources were classified using a color-color diagram.
There are 57 sources in group Ar,
reddened sources with little infrared excess.

\item
Aperture polarimetry of the point-like sources
resulted in the positive detection of 63 sources
in at least one of the three bands.
Most of the near-IR polarizations of the point-like sources
can be explained by dichroic polarization.

\item
Toward the HH 1--2 region,
L1641 is the only molecular cloud that is optically thick enough to cause the dichroic extinction.
For the group Ar sources,
the polarization direction does not depend on the amount of extinction,
which suggests that the L1641 cloud
is the only source of systematic polarization.

\item
Sources in group Ar are expected to be
either background stars or PMS stars with little intrinsic polarization.
The histogram of polarization position angles of group Ar sources
has a well-defined peak at $\sim$111\arcdeg,
which we interpret as the projected direction of the magnetic fields
in the HH 1--2 region.
From the 11\arcdeg\ dispersion of the polarization angles,
a rough estimate of the strength of the magnetic field
projected on the plane of the sky
is $\sim$130 $\mu$G.

\item
The orientation of the YSO outflows/jets in the HH 1--2 region
appears to be almost random,
which is completely different
from the distribution of the magnetic field directions in the cloud.
This difference suggests
that protostars may be disoriented during the star formation process,
probably because of the dynamical interaction in multiple systems.

\item
For the group Ar sources, the wavelength dependence of polarization
is consistent with the dichroic extinction.
Sources in group A0 have a small amount of extinction,
and their polarization seems to be caused by the circumstellar scattering.

\end{enumerate}

\acknowledgments

This work was supported
by the Korea Science and Engineering Foundation (KOSEF)
grant No. 2009-0063616, funded by the Korean government (MEST).
M.T. and T.N. are supported by a Grant-in-Aid from the Ministry
of Education, Culture, Sports, Science and Technology (No. 19204018).
This paper uses observations
made at the South African Astronomical Observatory.
IRAF is distributed by the US National Optical Astronomy Observatories,
which are operated by the Association of Universities
for Research in Astronomy, Inc.,
under cooperative agreement with the National Science Foundation.
This publication makes use of data products
from the Two Micron All Sky Survey,
which is a joint project of the University of Massachusetts
and the Infrared Processing
and Analysis Center/California Institute of Technology,
funded by the National Aeronautics and Space Administration
and the National Science Foundation.

\clearpage

\begin{deluxetable}{lcccccc}
\tabletypesize{\small}
\tablewidth{0pt}
\tablecaption{Photometry of Point-like Sources in the HH 1--2 Field}
\tablehead{
\colhead{Source} & \multicolumn{2}{c}{Position}
& \colhead{$J$} & \colhead{$H$} & \colhead{$K_s$}
& \colhead{Group\tablenotemark{a}} \\
\cline{2-3}
& \colhead {$\alpha_{\rm J2000.0}$} & \colhead {$\delta_{\rm J2000.0}$}
& \colhead{(mag)} & \colhead{(mag)} & \colhead{(mag)} & }%
\startdata
1  & 5 36 24.59 & --6 49 11.6 & 17.86 & 16.67 & 16.07 & Ar \\
2  & 5 36 27.99 & --6 49 11.1 & 17.32 & 15.99 & 15.34 & Ar \\
3  & 5 36 36.22 & --6 49 10.5 & 18.62 & 17.46 & 16.63 & B  \\
4  & 5 36 34.15 & --6 49 03.5 & 13.37 & 12.24 & 11.73 & Ar \\
5  & 5 36 32.60 & --6 49 00.1 & 18.08 & 16.81 & 16.30 & Ar \\
6  & 5 36 30.50 & --6 48 53.5 & 19.29 & 17.21 & 16.28 & Ar \\
7  & 5 36 35.41 & --6 48 44.6 & 14.77 & 13.85 & 13.41 & Ar \\
8  & 5 36 28.37 & --6 48 44.5 & 12.65 & 12.15 & 12.04 & A0 \\
9  & 5 36 26.35 & --6 48 43.4 & 13.58 & 13.09 & 12.73 & A0 \\
10 & 5 36 08.29 & --6 48 36.2 & 13.18 & 11.95 & 11.00 & B  \\
11 & 5 36 36.32 & --6 48 33.3 & 16.59 & 15.62 & 15.17 & Ar \\
12 & 5 36 37.65 & --6 48 22.5 & 16.29 & 14.96 & 14.40 & Ar \\
13 & 5 36 35.62 & --6 48 20.1 & 18.61 & 17.22 & 16.47 & Ar \\
14 & 5 36 37.57 & --6 48 13.2 & 16.74 & 15.68 & 15.20 & Ar \\
15 & 5 36 36.92 & --6 47 44.4 & 16.26 & 14.96 & 14.41 & Ar \\
16 & 5 36 35.58 & --6 47 18.4 & 18.09 & 16.46 & 15.67 & Ar \\
17 & 5 36 30.54 & --6 47 12.3 & 17.80 & 15.41 & 14.26 & Ar \\
18 & 5 36 37.46 & --6 46 57.5 & 16.10 & 14.78 & 14.34 & Ar \\
19 & 5 36 10.68 & --6 46 54.3 & 19.44 & 17.67 & 16.86 & Ar \\
20 & 5 36 32.15 & --6 46 46.0 & 18.44 & 16.73 & 16.00 & Ar \\
21 & 5 36 19.80 & --6 46 00.7 & 16.44 & 14.41 & 13.23 & B  \\
22 & 5 36 37.73 & --6 45 54.2 & 16.66 & 15.69 & 15.33 & Ar \\
23 & 5 36 15.70 & --6 45 53.1 & 15.79 & 15.28 & 15.04 & A0 \\
24 & 5 36 07.34 & --6 45 50.0 & 17.65 & 15.47 & 14.45 & Ar \\
25 & 5 36 29.62 & --6 45 48.2 & 16.51 & 14.02 & 12.87 & Ar \\
26 & 5 36 11.19 & --6 45 44.5 & 14.51 & 13.88 & 13.63 & A0 \\
27 & 5 36 30.71 & --6 45 38.5 & 14.93 & 12.14 & 10.76 & Ar \\
28 & 5 36 33.95 & --6 45 27.5 & 17.46 & 15.76 & 15.09 & Ar \\
29 & 5 36 23.95 & --6 45 23.8 & 13.16 & 12.56 & 12.29 & A0 \\
30 & 5 36 09.96 & --6 45 08.1 & 15.12 & 14.52 & 14.26 & A0 \\
31 & 5 36 38.06 & --6 45 08.4 & 16.87 & 15.71 & 15.31 & Ar \\
32 & 5 36 09.32 & --6 45 02.0 & 17.94 & 15.64 & 14.61 & Ar \\
33 & 5 36 31.84 & --6 44 47.2 & 18.37 & 16.37 & 15.54 & Ar \\
34 & 5 36 32.56 & --6 44 41.7 & 16.28 & 14.77 & 14.08 & Ar \\
35 & 5 36 28.10 & --6 44 32.5 & 13.20 & 12.22 & 11.74 & Ar \\
36 & 5 36 33.07 & --6 44 29.4 & 13.93 & 12.43 & 11.68 & Ar \\
37 & 5 36 23.58 & --6 44 27.0 & 17.17 & 15.28 & 14.03 & B  \\
38 & 5 36 12.10 & --6 44 23.3 & 16.78 & 14.77 & 13.85 & Ar \\
39 & 5 36 34.49 & --6 44 21.4 & 17.06 & 16.01 & 15.52 & Ar \\
40 & 5 36 32.88 & --6 44 20.9 & 12.86 & 11.32 & 10.55 & Ar \\
41 & 5 36 19.54 & --6 44 14.9 & 13.08 & 12.53 & 12.24 & A0 \\
42 & 5 36 32.09 & --6 44 14.2 & 17.96 & 16.46 & 15.78 & Ar \\
43 & 5 36 09.81 & --6 44 09.3 & 17.39 & 15.86 & 15.14 & Ar \\
44 & 5 36 07.19 & --6 44 08.8 & 19.04 & 16.90 & 15.91 & Ar \\
45 & 5 36 36.38 & --6 44 08.1 & 18.99 & 17.84 & 16.95 & B  \\
46 & 5 36 13.28 & --6 44 02.4 & 18.08 & 16.14 & 15.18 & Ar \\
47 & 5 36 11.37 & --6 44 00.1 & 16.69 & 15.23 & 14.55 & Ar \\
48 & 5 36 13.45 & --6 43 54.7 & 18.00 & 16.88 & 16.22 & B  \\
49 & 5 36 29.06 & --6 43 51.5 & 17.83 & 15.98 & 15.11 & Ar \\
50 & 5 36 26.77 & --6 43 43.4 & 13.84 & 11.94 & 11.07 & Ar \\
51 & 5 36 12.69 & --6 43 34.0 & 15.81 & 14.35 & 13.64 & Ar \\
52 & 5 36 15.27 & --6 43 30.8 & 19.22 & 17.02 & 16.01 & Ar \\
53 & 5 36 36.63 & --6 43 23.2 & 15.72 & 14.42 & 13.88 & Ar \\
54 & 5 36 34.28 & --6 43 23.2 & 13.17 & 12.51 & 12.33 & A0 \\
55 & 5 36 27.59 & --6 43 22.2 & 18.54 & 16.69 & 15.92 & Ar \\
56 & 5 36 30.18 & --6 43 20.4 & 17.19 & 15.77 & 15.14 & Ar \\
57 & 5 36 20.52 & --6 43 18.1 & 16.91 & 16.39 & 16.10 & A0 \\
58 & 5 36 07.34 & --6 43 07.6 & 12.89 & 11.88 & 11.42 & Ar \\
59 & 5 36 35.42 & --6 43 07.5 & 18.69 & 17.16 & 16.50 & Ar \\
60 & 5 36 25.94 & --6 43 02.2 & 13.85 & 13.01 & 12.55 & Ar \\
61 & 5 36 35.76 & --6 42 49.9 & 13.48 & 12.38 & 11.93 & Ar \\
62 & 5 36 36.87 & --6 42 49.4 & 19.10 & 17.02 & 16.07 & Ar \\
63 & 5 36 07.61 & --6 42 46.6 & 18.14 & 16.48 & 15.73 & Ar \\
64 & 5 36 30.23 & --6 42 46.1 & 13.31 & 11.81 & 11.09 & Ar \\
65 & 5 36 12.83 & --6 42 34.6 & 19.06 & 17.26 & 16.30 & Ar \\
66 & 5 36 35.14 & --6 42 18.6 & 18.25 & 16.66 & 16.00 & Ar \\
67 & 5 36 30.53 & --6 42 03.1 & 14.99 & 14.44 & 14.16 & A0 \\
68 & 5 36 20.82 & --6 41 56.3 & 19.52 & 17.53 & 16.71 & Ar \\
69 & 5 36 32.13 & --6 41 51.6 & 16.78 & 15.54 & 15.03 & Ar \\
70 & 5 36 17.49 & --6 41 46.1 & 18.28 & 17.08 & 16.42 & Ar \\
71 & 5 36 21.96 & --6 41 42.0 & 12.81 & 12.22 & 11.93 & A0 \\
72 & 5 36 35.81 & --6 41 41.3 & 16.95 & 16.38 & 16.05 & A0 \\
73 & 5 36 18.80 & --6 41 28.9 & 18.06 & 16.16 & 15.30 & Ar \\
74 & 5 36 35.37 & --6 41 29.2 & 16.08 & 14.84 & 14.33 & Ar \\
75 & 5 36 19.23 & --6 41 18.2 & 16.69 & 15.16 & 14.46 & Ar \\
76 & 5 36 31.52 & --6 41 13.6 & 12.72 & 11.98 & 11.78 & A0 \\
\enddata
\tablecomments{Units of right ascension are hours, minutes, and seconds,
               and units of declination are degrees, arcminutes,
               and arcseconds.
               Positions are from the $J$-$H$-$K_s$ image (Fig. 1).}
\tablenotetext{a}{Classification based on a color-color diagram
                  (see Section 4.2).}
\end{deluxetable}

\begin{deluxetable}{lrrrrrr}
\tabletypesize{\scriptsize}
\tablewidth{0pt}
\tablecaption{Polarimetry of Point-like Sources in the HH 1--2 Field}
\tablehead{
\colhead{Source} & \colhead{$P_J$} & \colhead{$P_H$} & \colhead{$P_{K_s}$}
& \colhead{$\theta_J$} & \colhead{$\theta_H$} & \colhead{$\theta_{K_s}$} \\
& \colhead{(\%)} & \colhead{(\%)} & \colhead{(\%)}
& \colhead{($\arcdeg$)} & \colhead{($\arcdeg$)} & \colhead{($\arcdeg$)} }%
\startdata
1 &   \mcoc{$<$ 12.2}    & \mcoc{$<$ \phn6.2}             & \mcoc{$<$ 20.4\phn}    & \mcoc{\ldots}        & \mcoc{\ldots}        & \mcoc{\ldots} \\
2 &   \mcoc{$<$ \phn9.6} & 3.6 \phn   $\pm$ 1.0 \phn\phn  & \mcoc{$<$ \phn9.5\phn} & \mcoc{\ldots}       & \phn 113.6 $\pm$ 8.0 & \mcoc{\ldots} \\
3 &   40.5 \phn $\pm$ \phn 8.5 \phn\phn & \mcoc{$<$ 10.8} & \mcoc{$<$ 32.8\phn}    & \phn 109.2 $\pm$ 5.9 & \mcoc{\ldots}        & \mcoc{\ldots} \\
4 &   \phn 2.74 $\pm$ \phn 0.08 \phn & 2.27 $\pm$ 0.04 \phn & \phn 1.55  $\pm$ 0.09 \phn & \phn\phn 97.7 $\pm$ 0.8 & \phn\phn 95.7 $\pm$    0.5 & \phn\phn 83.3 $\pm$ 1.6 \\
5 &   \mcoc{$<$ \phn9.4} & \mcoc{$<$ \phn4.4} & \mcoc{$<$ 14.4\phn} & \mcoc{\ldots} & \mcoc{\ldots} & \mcoc{\ldots} \\
6 &   \phn 31.3 \phn $\pm$ 10.2 \phn\phn & 5.7 \phn $\pm$ 1.8 \phn\phn & \mcoc{$<$ 12.7\phn} & \phn\phn 51.1 $\pm$ 8.9 & \phn 111.8 $\pm$ 8.5 & \mcoc{\ldots}           \\
7 &   \phn 2.84 $\pm$ \phn 0.17 \phn & 2.23 $\pm$ 0.09 \phn & \phn 1.1  \phn $\pm$ 0.3 \phn\phn & \phn 121.3 $\pm$ 1.7 & \phn 121.4 $\pm$ 1.1 & \phn   110.8 $\pm$   7.0 \\
8 &   \phn 0.30 $\pm$ \phn 0.04 \phn & 0.44 $\pm$ 0.03 \phn & \phn 0.42  $\pm$ 0.08 \phn & \phn 108.7 $\pm$ 3.7 & \phn 120.8 $\pm$ 2.0 & \phn\phn 94.9 $\pm$ 5.4 \\
9 &   \phn 0.33 $\pm$ \phn 0.07 \phn & 0.70 $\pm$ 0.05 \phn & \mcoc{$<$ \phn0.5\phn} & \phn 116.2 $\pm$ 6.3 & \phn 109.7 $\pm$ 2.1 &    \mcoc{\ldots} \\
10 & 10.97    $\pm$ \phn 0.06 \phn & 9.40 $\pm$ 0.03 \phn & \phn 7.30  $\pm$   0.04 \phn & \phn 148.6 $\pm$ 0.1 & \phn 147.7 $\pm$ 0.1 & \phn 147.9 $\pm$ 0.1 \\
11 & \phn 4.3 \phn $\pm$ \phn 0.7 \phn\phn & 3.4 \phn $\pm$ 0.3 \phn\phn & \mcoc{$<$ \phn4.3\phn} & \phn 143.6 $\pm$ 4.8 & \phn   127.9 $\pm$   2.9 &       \mcoc{\ldots} \\
12 & \phn 2.4 \phn $\pm$ \phn 0.6 \phn\phn & 2.2 \phn $\pm$ 0.3 \phn\phn & \mcoc{$<$ \phn2.7\phn} & \phn 119.8 $\pm$ 7.1 & \phn 116.0 $\pm$ 3.6 &       \mcoc{\ldots} \\
13 & \mcoc{$<$ 13.6}& \mcoc{$<$ \phn5.0} & \mcoc{$<$ 13.2\phn} & \mcoc{\ldots} & \mcoc{\ldots} & \mcoc{\ldots} \\
14 & \phn 6.2 \phn $\pm$ \phn 0.9 \phn\phn & 4.1 \phn $\pm$   0.5 \phn\phn & \mcoc{$<$ \phn5.2\phn} & \phn 118.6 $\pm$ 4.0 & \phn 116.9 $\pm$ 3.6 &       \mcoc{\ldots} \\
15 & \phn 1.8 \phn $\pm$ \phn 0.6 \phn\phn & 2.1 \phn $\pm$   0.2 \phn\phn & \mcoc{$<$ \phn2.1\phn} & \phn 122.2 $\pm$ 8.8 & \phn   125.3 $\pm$   3.2 &       \mcoc{\ldots} \\
16 & \mcoc{$<$ \phn8.4}& 2.9 \phn $\pm$ 0.8 \phn\phn & \mcoc{$<$ \phn5.9\phn} & \mcoc{\ldots} & \phn 121.7 $\pm$ 7.7 & \mcoc{\ldots} \\
17 & \phn 7.2 \phn $\pm$ \phn 2.1 \phn\phn & 3.1 \phn $\pm$ 0.3 \phn\phn & \phn 2.8 \phn $\pm$ 0.6 \phn\phn & \phn 127.8 $\pm$ 7.9 & \phn 133.5 $\pm$ 3.2 & \phn 119.8 $\pm$ 5.6 \\
18 & \phn 4.5 \phn $\pm$ \phn 0.5 \phn\phn & 3.3 \phn $\pm$ 0.2 \phn\phn & \phn 3.2 \phn $\pm$ 0.7 \phn\phn & \phn 115.5 $\pm$ 3.0 & \phn 118.7 $\pm$ 2.1 & \phn 123.8 $\pm$ 6.3 \\
19 & \mcoc{$<$ 30.2} & \mcoc{$<$ \phn7.0} & \mcoc{$<$ 19.2\phn} & \mcoc{\ldots} & \mcoc{\ldots} & \mcoc{\ldots} \\
20 & \mcoc{$<$ 10.2} & 7.3 \phn   $\pm$ 1.2 \phn\phn & \mcoc{$<$ \phn8.8\phn} & \mcoc{\ldots} & \phn 132.1 $\pm$ 4.7 & \mcoc{\ldots} \\
21 & \phn 7.7 \phn $\pm$ \phn 0.6 \phn\phn & 6.07 $\pm$   0.15 \phn & \phn 3.9 \phn $\pm$ 0.2 \phn\phn & \phn\phn 20.3 $\pm$ 2.2 & \phn\phn 21.0 $\pm$ 0.7 & \phn\phn   14.0 $\pm$   1.7 \\
22 & \mcoc{$<$ \phn2.7}   & 2.8 \phn $\pm$ 0.5 \phn\phn & \mcoc{$<$ \phn6.4\phn} & \mcoc{\ldots} & \phn 118.2 $\pm$ 5.3 & \mcoc{\ldots} \\
23 & \mcoc{$<$ \phn1.2}   & \mcoc{$<$ \phn0.9} & \mcoc{$<$ \phn3.4\phn} & \mcoc{\ldots} & \mcoc{\ldots} & \mcoc{\ldots} \\
24 & \mcoc{$<$ \phn8.3}   & 4.4 \phn   $\pm$ 0.4 \phn\phn & \phn 5.0 \phn   $\pm$ 0.8 \phn\phn & \mcoc{\ldots} & \phn\phn 84.0 $\pm$ 2.6 & \phn\phn 82.6 $\pm$    4.5 \\
25 & \phn 6.9 \phn   $\pm$ \phn 0.7 \phn\phn & 5.00 $\pm$ 0.10 \phn & \phn 3.21  $\pm$ 0.16 \phn & \phn 100.3 $\pm$ 2.9 & \phn 106.6 $\pm$ 0.6 &   \phn 108.5 $\pm$ 1.4 \\
26 & \phn 0.67 $\pm$ \phn 0.17 \phn & 0.96 $\pm$ 0.12 \phn & \mcoc{$<$ \phn1.1\phn} & \phn 110.0 $\pm$   7.1 & \phn 126.0 $\pm$ 3.6 & \mcoc{\ldots} \\
27 & \phn 5.89 $\pm$ \phn 0.18 \phn & 4.03 $\pm$ 0.03 \phn & \phn 2.59  $\pm$ 0.03 \phn & \phn 114.6 $\pm$ 0.9 & \phn 113.5 $\pm$   0.2 & \phn 112.5 $\pm$ 0.3 \\
28 & \phn 7.3 \phn $\pm$ \phn 1.6 \phn\phn & 4.5 \phn $\pm$ 0.5 \phn\phn & \mcoc{$<$ \phn3.5\phn} & \phn 118.9 $\pm$ 5.9 & \phn 127.4 $\pm$   2.9 & \mcoc{\ldots} \\
29 & \phn 0.51 $\pm$ \phn 0.05 \phn & 0.74 $\pm$ 0.04 \phn & \phn 0.53 $\pm$ 0.09 \phn & \phn 118.9 $\pm$ 3.0 & \phn 112.8 $\pm$   1.5 & \phn\phn 99.6 $\pm$ 5.0 \\
30 & \phn 0.62 $\pm$ \phn 0.20 \phn & 0.65 $\pm$ 0.16 \phn & \phn 2.4 \phn $\pm$ 0.6 \phn\phn & \phn 107.3 $\pm$ 8.7 & \phn 123.3    $\pm$ 6.7 & \phn\phn 65.8 $\pm$ 6.6 \\
31 & \phn 4.8 \phn $\pm$ \phn 1.3 \phn\phn & \mcoc{$<$ \phn2.0} & \phn 7.8 \phn $\pm$ 2.3 \phn\phn & \phn 115.5 $\pm$ 7.5 & \mcoc{\ldots} &   \phn 137.9 $\pm$ 8.1 \\
32 & \phn 13.6 \phn   $\pm$ \phn 2.5 \phn\phn   & 5.2 \phn $\pm$ 0.4 \phn\phn & \phn 3.5 \phn $\pm$ 0.8 \phn\phn & \phn\phn 79.0 $\pm$   5.1 & \phn\phn 82.6 $\pm$ 2.2 & \phn\phn 72.3 $\pm$ 6.3 \\
33 & \mcoc{$<$ \phn9.6}   & 5.9 \phn $\pm$ 0.9 \phn\phn & \mcoc{$<$ \phn5.6\phn} & \mcoc{\ldots} & \phn 114.2 $\pm$ 4.4 & \mcoc{\ldots} \\
34 & \phn 6.5 \phn   $\pm$ \phn 0.5 \phn\phn   & 5.2 \phn $\pm$ 0.2 \phn\phn &   \phn 4.1 \phn $\pm$ 0.5 \phn\phn & \phn 126.0 $\pm$ 2.2 & \phn 115.1 $\pm$ 1.1 & \phn   130.4 $\pm$ 3.3 \\
35 & \phn 0.71 $\pm$ \phn 0.05 \phn & 0.83 $\pm$ 0.03 \phn & \phn 0.65  $\pm$ 0.06 \phn & \phn\phn 71.6 $\pm$ 2.1 & \phn\phn 71.3 $\pm$ 1.1 & \phn\phn 70.9 $\pm$ 2.8 \\
36 & \phn 1.49 $\pm$ \phn 0.10 \phn & 1.12 $\pm$ 0.04 \phn & \phn 0.68  $\pm$ 0.08 \phn & \phn 118.0 $\pm$ 2.0 & \phn 109.5 $\pm$ 1.1 & \phn 109.1 $\pm$ 3.3 \\
37 & \mcoc{$<$ \phn3.5} & \mcoc{$<$ \phn1.0} & \phn   1.7 \phn $\pm$ 0.5 \phn\phn & \mcoc{\ldots} & \mcoc{\ldots} & \phn 118.8 $\pm$   8.1 \\
38 & \phn 2.9 \phn $\pm$ \phn 0.8 \phn\phn & 2.3 \phn $\pm$ 0.2 \phn\phn & \phn 1.8 \phn $\pm$ 0.4 \phn\phn & \phn 124.5 $\pm$ 8.1 & \phn 111.6 $\pm$ 2.7 & \phn 109.0 $\pm$ 6.3 \\
39 & \phn 3.9 \phn $\pm$ \phn 1.0 \phn\phn & 3.2 \phn $\pm$ 0.5 \phn\phn & \mcoc{$<$ \phn5.7\phn} & \phn 107.8 $\pm$ 6.7 & \phn 113.1 $\pm$ 4.5 & \mcoc{\ldots} \\
40 & \phn 1.59 $\pm$ \phn 0.05 \phn & 1.38 $\pm$ 0.03 \phn & \phn 0.93  $\pm$ 0.03 \phn & \phn 132.8 $\pm$ 0.9 & \phn 128.0 $\pm$ 0.5 & \phn 126.7 $\pm$ 0.9 \\
41 & \phn 0.57 $\pm$ \phn 0.05 \phn & 0.65 $\pm$ 0.04 \phn & \phn 0.48  $\pm$ 0.09 \phn & \phn 115.9 $\pm$ 2.7 & \phn 114.8 $\pm$ 1.6 & \phn 112.8 $\pm$ 5.4 \\
42 & \mcoc{$<$ \phn8.2}   & 5.0 \phn $\pm$ 1.2 \phn\phn &   \mcoc{$<$ \phn7.5\phn} & \mcoc{\ldots} & \phn 114.7 $\pm$ 6.4 & \mcoc{\ldots} \\
43 & \mcoc{$<$ \phn4.3} & 2.4 \phn $\pm$ 0.5 \phn\phn & \mcoc{$<$ \phn3.8\phn} & \mcoc{\ldots} & \phn 107.8 $\pm$ 5.3 & \mcoc{\ldots} \\
44 & \mcoc{$<$    26.1} & 5.6 \phn $\pm$ 1.7 \phn\phn & \mcoc{$<$ \phn9.5\phn} & \mcoc{\ldots} & \phn 126.2 $\pm$ 8.2 &   \mcoc{\ldots} \\
45 & \mcoc{$<$    15.0} & \mcoc{$<$ 10.0}             & \mcoc{$<$    20.7\phn} & \mcoc{\ldots} & \mcoc{\ldots}         &   \mcoc{\ldots} \\
46 & 10.9 \phn $\pm$ \phn 2.5 \phn\phn & 5.4 \phn $\pm$   0.7 \phn\phn & \mcoc{$<$ \phn4.1\phn} &   \phn\phn 82.9 $\pm$   6.3 & \phn 107.1 $\pm$ 3.9 &       \mcoc{\ldots} \\
47 & \phn 3.5 \phn $\pm$ \phn 0.7 \phn\phn & 2.0 \phn $\pm$ 0.3 \phn\phn & \mcoc{$<$ \phn2.3\phn} & \phn\phn 84.2 $\pm$ 5.9 & \phn 113.6 $\pm$ 3.7 &       \mcoc{\ldots} \\
48 &     17.8 \phn $\pm$ \phn 2.5 \phn\phn & 8.8 \phn $\pm$ 1.3 \phn\phn & \mcoc{$<$    10.1\phn} & \phn    149.5 $\pm$ 4.0 & \phn 133.9 $\pm$ 4.1 &       \mcoc{\ldots} \\
49 & \mcoc{$<$ \phn7.5}   & 5.2 \phn $\pm$ 0.6 \phn\phn &   \mcoc{$<$ \phn3.8\phn} & \mcoc{\ldots} & \phn 104.6 $\pm$ 3.2 &   \mcoc{\ldots} \\
50 & \phn 1.64  $\pm$ \phn 0.11 \phn & 1.09 $\pm$ 0.04 \phn & \phn 0.56  $\pm$ 0.04 \phn & \phn\phn 90.3 $\pm$ 1.8 & \phn\phn 91.8 $\pm$ 1.0 & \phn 109.0 $\pm$ 2.0 \\
51 & \phn 4.7 \phn $\pm$ \phn 0.4 \phn\phn & 3.03 $\pm$   0.14 \phn & \phn 1.8 \phn $\pm$   0.3 \phn\phn & \phn\phn 86.9 $\pm$ 2.1 & \phn\phn 95.3 $\pm$   1.3 & \phn\phn 99.2 $\pm$ 4.9 \\
52 & \mcoc{$<$ 21.9} & \mcoc{$<$ \phn4.2} & \mcoc{$<$ \phn8.7\phn} & \mcoc{\ldots} & \mcoc{\ldots} & \mcoc{\ldots} \\
53 & \phn 3.6 \phn $\pm$ \phn 0.3 \phn\phn & 2.56 $\pm$ 0.16 \phn & \phn 1.7 \phn $\pm$ 0.4 \phn\phn & \phn 131.5 $\pm$ 2.7 & \phn 122.8 $\pm$ 1.8 & \phn 137.9 $\pm$ 6.5 \\
54 & \phn 0.66    $\pm$ \phn 0.05 \phn & 0.70 $\pm$ 0.04 \phn & \phn 0.63 $\pm$ 0.10 \phn & \phn 125.6 $\pm$ 2.4 & \phn 120.7 $\pm$ 1.4 & \phn 110.7 $\pm$ 4.6 \\
55 & \mcoc{$<$ 17.8} & 8.3 \phn $\pm$ 1.4 \phn\phn & 10.3 \phn $\pm$ 2.8 \phn\phn & \mcoc{\ldots} & \phn 106.4 $\pm$ 4.8 & \phn 128.4 $\pm$ 7.6 \\
56 & \mcoc{$<$ \phn3.7}   & 3.4 \phn $\pm$ 0.5 \phn\phn &   \mcoc{$<$ \phn3.7\phn} & \mcoc{\ldots} & \phn 110.1 $\pm$ 4.1 & \mcoc{\ldots} \\
57 & \mcoc{$<$ \phn3.1}   & 2.6 \phn $\pm$ 0.8 \phn\phn &   \mcoc{$<$ \phn9.4\phn} & \mcoc{\ldots} & \phn 104.5 $\pm$ 8.1 & \mcoc{\ldots} \\
58 & \phn 4.14  $\pm$ \phn 0.05 \phn & 2.54 $\pm$ 0.03 \phn & \phn 1.39  $\pm$ 0.06 \phn & \phn\phn 92.9 $\pm$ 0.3 & \phn\phn 95.2 $\pm$ 0.3 & \phn 103.5 $\pm$ 1.2 \\
59 & \mcoc{$<$    13.7} & \mcoc{$<$ \phn4.9} & \mcoc{$<$ 13.5\phn}    & \mcoc{\ldots} & \mcoc{\ldots} & \mcoc{\ldots} \\
60 & \mcoc{$<$ \phn3.6}   & \mcoc{$<$ \phn2.8} & \mcoc{$<$ \phn2.5\phn} & \mcoc{\ldots} & \mcoc{\ldots} & \mcoc{\ldots} \\
61 & \phn 2.31  $\pm$ \phn 0.07 \phn & 1.70 $\pm$ 0.04 \phn & \phn 1.08  $\pm$ 0.07 \phn & \phn 130.5 $\pm$ 0.8 & \phn 125.4 $\pm$ 0.6 & \phn 114.5 $\pm$ 1.9 \\
62 & \mcoc{$<$ 17.7} & 7.0 \phn $\pm$ 1.5 \phn\phn & \mcoc{$<$ \phn9.8\phn} & \mcoc{\ldots} & \phn 135.6 $\pm$ 5.9 & \mcoc{\ldots} \\
63 & \mcoc{$<$ 10.6} & \mcoc{$<$ \phn2.8} & \mcoc{$<$ \phn6.6\phn} & \mcoc{\ldots} & \mcoc{\ldots} & \mcoc{\ldots} \\
64 & \phn 1.47 $\pm$ \phn 0.06 \phn & 1.21 $\pm$ 0.03 \phn & \phn 0.81 $\pm$ 0.04 \phn & \phn 126.7 $\pm$ 1.1 & \phn 129.3 $\pm$ 0.6 & \phn 138.5 $\pm$ 1.3 \\
65 & \mcoc{$<$ 17.3} & \mcoc{$<$ \phn4.7} &   \mcoc{$<$ 11.3\phn}   & \mcoc{\ldots} & \mcoc{\ldots} & \mcoc{\ldots} \\
66 & 13.7 \phn $\pm$ \phn 3.1 \phn\phn & \mcoc{$<$ \phn3.8} & \mcoc{$<$ \phn8.8\phn} & \phn 120.3 $\pm$ 6.3 & \mcoc{\ldots} & \mcoc{\ldots} \\
67 & \mcoc{$<$ \phn0.5} & 0.52 $\pm$ 0.15 \phn & \phn 2.0 \phn $\pm$ 0.5 \phn\phn & \mcoc{\ldots} & \phn\phn 98.6 $\pm$ 8.1 & \phn 124.5 $\pm$ 7.3 \\
68 & \mcoc{$<$ 33.9} & \mcoc{$<$ \phn7.5} &   \mcoc{$<$ 16.2\phn} & \mcoc{\ldots}   & \mcoc{\ldots}   & \mcoc{\ldots} \\
69 & \phn 7.2 \phn $\pm$ \phn 0.8 \phn\phn & 4.3 \phn $\pm$ 0.4 \phn\phn & \mcoc{$<$ \phn3.4\phn} & \phn 115.7 $\pm$ 3.2 & \phn 118.6 $\pm$ 2.6 & \mcoc{\ldots} \\
70 & \phn 9.9 \phn $\pm$ \phn 3.1 \phn\phn & \mcoc{$<$ \phn4.6} & 13.1 \phn $\pm$ 4.3 \phn\phn & \phn\phn 35.6 $\pm$ 8.6 & \mcoc{\ldots} & \phn 100.9 $\pm$ 8.9 \\
71 & \phn 0.43 $\pm$ \phn 0.05 \phn & 0.62 $\pm$ 0.03 \phn & \phn 0.45 $\pm$ 0.07 \phn & \phn 135.8 $\pm$ 3.0 & \phn 134.6 $\pm$ 1.4 & \phn 134.3 $\pm$ 4.6 \\
72 & \mcoc{$<$ \phn4.0}   & \mcoc{$<$ \phn2.5} & \mcoc{$<$ \phn9.2\phn} & \mcoc{\ldots} & \mcoc{\ldots} & \mcoc{\ldots} \\
73 & \mcoc{$<$ 10.0} & 5.1 \phn   $\pm$ 0.8 \phn\phn & \mcoc{$<$ \phn5.2\phn} & \mcoc{\ldots} & \phn 101.3 $\pm$ 4.2 & \mcoc{\ldots} \\
74 & \phn 2.7 \phn $\pm$ \phn 0.7 \phn\phn & 3.5 \phn $\pm$ 0.2 \phn\phn & \mcoc{$<$ \phn2.2\phn} & \phn 120.4 $\pm$ 7.5 & \phn 120.9 $\pm$ 2.0 & \mcoc{\ldots} \\
75 & \phn 8.8 \phn $\pm$ \phn 1.4 \phn\phn & 4.6 \phn $\pm$ 0.4 \phn\phn & \phn   4.6 \phn $\pm$ 1.0 \phn\phn & \phn\phn 88.1 $\pm$ 4.3 & \phn 101.0 $\pm$ 2.6 & \phn 123.3 $\pm$ 6.3 \\
76 & \phn 0.58 $\pm$ \phn 0.06 \phn & 0.53 $\pm$ 0.03 \phn & \mcoc{$<$ \phn0.34} & \phn 110.0 $\pm$ 2.8 & \phn 105.0 $\pm$ 1.8 & \mcoc{\ldots} \\
\enddata
\tablecomments{For sources with $P/\delta P < 3$, the 3$\delta P$ upper limits are listed.}
\end{deluxetable}

\begin{deluxetable}{lrcc}
\tabletypesize{\small}
\tablewidth{0pt}
\tablecaption{Optical Jets in the HH 1--2 Field}
\tablehead{
\colhead{Object} & \colhead{P.A.\tablenotemark{a}}
& \colhead{Driving Source} & \colhead{References}}
\startdata
HH 1--2     & 148\arcdeg & VLA 1      & Pravdo et al. 1985 \\
HH 35       & 149\arcdeg & V380 Ori   & Strom et al. 1986 \\
HH 144--145 &  82\arcdeg & VLA 2      & Reipurth et al. 1993 \\
HH 146      &   6\arcdeg & VLA 4      & Reipurth et al. 1993 \\
HH 147      &  50\arcdeg & N$^3$SK 50 & Eisl{\"o}ffel et al. 1994 \\
HH 148      &  56\arcdeg & V380 Ori   & Strom et al. 1986 \\
SMZ 61      & 171\arcdeg & VLA 3      & Stanke et al. 2002 \\
\enddata
\tablenotetext{a}{Position angle of the outflow axis.
                  Typical uncertainty is 10\arcdeg.}
\end{deluxetable}

\end{document}